# Achieving huge thermal conductance of metallic nitride on graphene through enhanced elastic and inelastic phonon transmission

Weidong Zheng[1,2], Bin Huang[1,2], Hongkun Li[1], Yee Kan Koh[1,2*]

[1]Department of Mechanical Engineering, National University of Singapore, Singapore 117576

[2]Centre for Advanced 2D Materials, National University of Singapore, Singapore 117542

*Corresponding Author. Email: mpekyk@nus.edu.sg

**Abstract**

Low thermal conductance of metal contacts is one of the main challenges in thermal management of nanoscale devices of graphene and other 2D materials. Previous attempts to search for metal contacts with high thermal conductance yielded limited success due to incomplete understanding of the origins of the low thermal conductance. In this paper, we carefully study the intrinsic thermal conductance across metal/graphene/metal interfaces to identify the heat transport mechanisms across graphene interfaces. We find that unlike metal contacts on diamond, the intrinsic thermal conductance of most graphene interfaces (except Ti and $TiN_x$) is only $\approx$ 50 % of the phonon radiation limit, suggesting that heat is carried across graphene interfaces mainly through elastic transmission of phonons. We thus propose a convenient approach to substantially enhance the phononic heat transport across metal contacts on graphene, by better matching the energy of phonons in metals and graphene, e.g., using metallic nitrides. We test the idea with $TiN_x$, with phonon frequencies of up to $1.18 \times 10^{14}$ rad/s, 47 % of the highest phonon frequencies in graphene of $2.51 \times 10^{14}$ rad/s . Interestingly, we obtain a huge thermal conductance of 270 MW m$^{-2}$ K$^{-1}$ for $TiN_x$/graphene interfaces, which is $\approx$ 140 % of the phonon radiation limit. The huge thermal conductance could be partially attributed to inelastic phonon transport across the $TiN_x$/graphene interface. Our work provide guidance for the search for good metal contacts on 2D materials and devices.

**Text**

Effective thermal management is a recurrent challenge for graphene-based devices, especially when the devices operate under a high electric field with a high current density.[1-5] Without efficient heat dissipation from the active regions, carrier mobility[3] and breakdown current density[5] of graphene could be suppressed, due to self-heating of the devices and formation of hot spots. While the in-plane thermal conductivity of graphene is exceptionally high ($\approx 600$ W m$^{-1}$ K$^{-1}$),[6] heat transfer in graphene devices is still severely impeded by graphene interfaces (i.e., substrate/graphene and metal/graphene interfaces), as graphene is only single atomic-layer thick. In fact, when the lateral size of graphene devices is sub-micrometer (e.g., $\leq$ 500 nm), heat dissipation from graphene devices is predominantly limited by heat conduction across metal contacts.[7] Therefore, it is vital to search for metal contacts with high thermal conductance, for efficient thermal management of nanoscale graphene devices.

Up to now, prior reported measurements of the thermal conductance of metal/graphene interfaces are relatively low. (The thermal conductance of most metal contacts[7-14] on graphene ranges 50 – 100 MW m$^{-2}$ K$^{-1}$, while the thermal conductance of other epitaxial solid/solid interfaces[15] could be as high as 700 MW m$^{-2}$ K$^{-1}$.) Previous attempts to enhance the low thermal conductance of metal/graphene interfaces have only yielded limited success. One possible route to enhance the thermal conductance of graphene interfaces is by altering the bonding strength of metals and graphene, through chemical functionalization of graphene by adsorbates.[12, 16] While an improvement of up to 50 % was reported, the reported thermal conductance of the functionalized graphene interfaces is still only < 50 MW m$^{-2}$ K$^{-1}$, far below the thermal conductance usually observed for other solid/solid interfaces. Another possible route to enhance the thermal conductance of graphene interfaces is by boosting the electronic heat transport across graphene. We previously demonstrated that charge carriers play a negligible role in heat transfer across intrinsic Pd/graphene/Pd interfaces, even when graphene is significantly doped ($\approx 3 \times 10^{12}$ cm$^{-2}$) by Pd.[13] We showed that the electronic heat transport across the metal/graphene interfaces is only substantially enhanced when atomic-scale pinholes were created in the graphene via magnetron sputtering or ion bombardment.[13] The damages induced in graphene, however, represent a serious drawback for the approach, because the performance and functionality of graphene devices could be affected by the atomic-scale pinholes.

One of the reasons contributing to the failure to find metals that make excellent thermal contacts with graphene is lack of in-depth knowledge of the intrinsic heat transport across graphene interfaces. For example, up to now, it is still unclear whether the low thermal conductance of graphene interfaces is due to a large disparity in phonon energy in graphene and metals,[7] or the bonding strength of metal/graphene interfaces (i.e., physisorption or chemisorption bonds).[12, 17-18] Also, the role of inelastic phonon scattering at interfaces, which is responsible for heat transport across metal/diamond interfaces[19-21], is yet to be determined. The incomplete understanding is partly a consequence of an insufficient number of systematic and careful experimental studies[7-8, 10, 12, 22] on the *intrinsic thermal conductance* of graphene interfaces, due to practical challenges in the preparation of intrinsic graphene interfaces.

Particularly, it is crucial to minimize the amounts of extrinsic defects (e.g., voids and polymer residues) at graphene interfaces, which are inevitably present after graphene transfer but could further impede the interfacial heat transport. Through our prior careful work on Al/graphene/Cu interfaces, we demonstrated that the thermal conductance of interfaces of transferred graphene only approaches the intrinsic values if graphene conforms to the substrates (no voids) and if the amount of polymer residues is sufficiently small.[10] Unfortunately, the critical information on the conformity of graphene and the amount of polymer residues (as evidenced by the high-resolution atomic force microscopy images[10]) is missing in most previous studies.[8-9, 11-12, 14, 22]

Moreover, in most prior studies, graphene was sandwiched between two different materials[7-10, 12]. While most researchers follow our previous work to derive the thermal conductance of individual interfaces,[7] by assuming that the thermal resistance of individual interfaces adds in series, there is an additional uncertainty with such approximation. Also, even for cases of graphene being sandwiched between the same metals,[11, 13-14] it is important to ensure that there is no native oxide on the bottom metals, because a thin layer of oxide could significantly affect the transmission of phonons and electrons across the interfaces.[13]

In this paper, we investigate the *intrinsic* heat transport across metal/graphene interfaces, by measuring the thermal conductance of interfaces of graphene sandwiched between the same metals that do not oxidize under atmospheric conditions. We carefully ensure that the transferred graphene conforms fully to the bottom metal layers, and the polymer residues after the graphene transfer are minimal. We compare the *intrinsic thermal conductance* (*G*) of metal/graphene interfaces with the phonon radiation limit ($G_{rad}$),[23] which only depends on the highest phonon energy in the metals. From the comparison, we find that heat transport across most graphene interfaces (except Ti and $TiN_x$) is mostly due to elastic phonon transmission across the interfaces, and thus conclude that the low intrinsic thermal conductance of graphene interfaces is predominantly due to a huge mismatch in phonon energy in graphene and previously explored metals. For metallic $TiN_x$, however, we achieve $G = 270$ MW m$^{-2}$ K$^{-1}$ for the $TiN_x$/graphene interface. The measured $G$ is larger than $G_{rad}$ of $TiN_x$/graphene interface, and thus cannot be explained by a better match of phonon energy alone. We attribute the high thermal conductance partially to the inelastic heat transport across the $TiN_x$ contact on graphene. Our results hence provide an important guide to search for metals with good thermal contacts with other 2D materials.

Our samples comprise of transferred graphene sandwiched between two layers of metal films on GaN/sapphire substrates, see Figure 1a. Details of the sample preparation are summarized in Methods. We carefully select three metals (i.e., Au, Ag and $TiN_x$) that do not oxidize under atmospheric conditions, to study the heat transport across intrinsic metal/graphene interfaces. We postulate that the low thermal conductance of previously reported graphene interfaces could be due to weak interfacial bonding of graphene and prior explored metals, and/or a large mismatch in the phonon energy in graphene and the metals, see the comparison of phonon density of states (DOS) of Ag, Au, $TiN_x$ and graphene in Figure 1b. Thus, we explore

TiN$_x$ as a suitable candidate for excellent thermal contacts, since TiN$_x$ forms chemisorption interfaces[24] with graphene and has a smaller mismatch in the phonon energy with graphene.

We ensure that our graphene is pristine and the interfaces are intrinsic by carefully characterizing our samples by Raman spectroscopy and atomic force microscopy (AFM). We apply Raman spectroscopy to examine the quality of the graphene after thermal or e-beam evaporation of metals, see Figure 1c. We observe no significant D peaks in all samples after metal evaporation, suggesting that the graphene is undamaged even after the thermal and e-beam evaporation.[7, 10] We employ tapping mode AFM to confirm that the graphene in our samples conforms to the substrate without excessive polymer residues. As shown in the topographic images of our graphene on metal films in Figure 1d, we find that the roughness of samples does not substantially change before and after graphene transfer; for the Au/G/Au sample, we obtained a low root-mean-square (rms) roughness of 0.96 nm after graphene transfer, similar to the rms roughness 0.86 nm before graphene transfer. We also obtained the AFM phase images of the graphene after the transfer and observe no distinct phase differences, confirming that the transfer is clean with little polymer residues, see Figure 1e. Moreover, using a method described in Methods and ref 10, we find that the contact area is approximately 100 % for all our samples, see Figure 1f, indicating that graphene conforms fully to the metal films.

We measure the thermal conductance ($G_{M/G/M}$) of metal/graphene/metal interfaces (M/G/M, where G denotes graphene and M denotes the metals) by time-domain thermoreflectance (TDTR); details of our implementation are discussed in Methods and our previous papers.[10, 25] We note that $G_{M/G/M}$ is the only fitting parameter in our measurements, and thus the derive $G_{M/G/M}$ are very reliable.

In Figure 2a, we report the measured $G_{M/G/M}$ over a temperature range of $80 \leq T \leq 300$ K. For comparison, we also include prior reported thermal conductance of graphene interfaces, including Au/G/SiO$_2$[7], Al/G/Cu[10], Al/G/SiO$_2$[12], Ti/G/ SiO$_2$[8], Pd/G/Pd[13], and Ti/G/Ti[14]. The comparison, however, could yield limited physical understanding on the intrinsic thermal transport across *single* graphene interfaces, considering the wide variety of the interface combinations in prior studies. Thus, we derive the thermal conductance of single metal/graphene interfaces, $G_{M/G}$, from our measurements in Figure 2a, by assuming that the thermal resistances of the top and bottom metal/G interfaces add in series;[7] $G_{M/G} = 2G_{M/G/M}$. We plot the derived $G_{M/G}$ in Figure 2b, and compare the values to $G_{M/G}$ derived from prior measurements of M/G/M interfaces[13-14] and thermal conductance of a wide range of interfaces of other 2D materials, including Al/graphite,[26] Ti/graphite,[26] Al/BP,[27] Al/MoS$_2$,[28] MoS$_2$/h-BN,[29] and graphene/h-BN[29] interfaces.

We find that for the Au/G/Au and Ag/G/Ag samples, the thermal conductances are relatively low and comparable to prior reported thermal conductance of interfaces of graphene[7-8, 10, 12, 14, 30] and MoS$_2$[29, 31], see Figures 2a and 2b. Interestingly, for TiN$_x$/G/TiN$_x$ samples, the thermal conductance is significantly higher than that of other graphene and MoS$_2$ interfaces. Specifically, at room temperature, $G_{\text{TiN}_x/G}$ = 270 MW m$^{-2}$ K$^{-1}$, two times larger than $G$ of most interfaces compiled in Figure 2b.

To gain more insights to the thermal transport mechanisms across graphene interfaces and thus understand the origins of the huge $G_{TiNx/G}$, we compile our measurements and prior reported thermal conductance measurements ($G_{exp}$) for a wide range of metal/graphene (blue symbols[13-14] for $G_{M/G}$ derived from $G_{M/G/M}$, black symbols[7-8, 12, 14, 16, 30] for $G_{M/G}$ derived from $G_{M/G/SiO2}$), metal/graphite (red symbols)[14, 26] and graphene/SiO$_2$ (purple symbol)[22] interfaces in Fgure 3. (To derive $G_{M/G}$ from prior measurements, we use $G_{M/G} = (G_{M/G/SiO_2}^{-1} - G_{G/SiO_2}^{-1})^{-1}$ and $G_{G/SiO2}$ = 97 MW m$^{-2}$ K$^{-1}$ from ref 22.) In the analysis, we also compare the compiled thermal conductance to the phonon radiation limit ($G_{rad}$) of the interfaces, plotting $G_{rad}$, 0.5×$G_{rad}$, and 0.25×$G_{rad}$ in Figure 3. The phonon radiation limit[19] is the highest interfacial thermal conductance in theory if only elastic phonon transmission is permissible, and it occurs when all phonons from the side with lower phonon irradiation transmit elastically across the interfaces. Details of the calculations of $G_{rad}$ are presented in Methods. We note that for graphene interfaces, $G_{rad}$ can be approximated by eq 3 and thus only depends on the highest phonon frequency in the metal, $\omega_{max}$.

We first notice in Figure 3a that $G_{exp}$ of some interfaces (e.g., for Pd, Ti and Al contacts) reported by different researchers spreads a wide range. For example, $G_{Ti/G}$ derived from measurements on Ti/G/SiO$_2$ interface by Schmidt et al.[30] is 7× smaller than that derived from measurements on the same Ti/G/SiO$_2$ interface by Goodson et al.[8] The discrepancy observed in Figure 3a demonstrates the importance of careful preparation of intrinsic interfaces and accurate thermal measurements. Additional thermal resistance could be measured if the graphene interfaces are not intrinsic, i.e., if graphene is contaminated with polymer residues or does not conform fully to the substrate.[10] Thus, we omit measurements that are substantially lower in the subsequent analysis (e.g., in Figure 3b and 3c).

Interestingly, we observe that for most compiled $G_{exp}$, the $G_{M/G}$ of graphene interfaces is mostly larger than $G$ of metal/graphite interfaces, see Figure 3b. The finding could be due to partial transmission of low-energy phonons across both top and bottom graphene interfaces, akin to transmission of low-energy phonons that are weakly scattered by interfaces in short-period AlN/GaN superlattices.[32] (For the AlN/GaN superlattices, the apparent thermal conductance increases more than threefold when the period is reduced from 40 nm to 2 nm.) In other words, while the top and bottom interfaces of single-atomic-layer thick graphene is largely decoupled as previously concluded,[7] the decoupling is not as complete as previously thought. In fact, careful re-examination of the previous measurements on the thickness-dependent thermal conductance of Au/G/SiO$_2$ in ref 7 indicates a ≈30 % difference in the thermal conductance of interfaces of single-layer graphene and 10-layered graphene, consistent with the conclusion found in this paper.

Finally, we consider three possible explanations to the enhanced heat transport across TiN$_x$/G interfaces. The interfacial heat transport could be enhanced by better matching phonons energy in TiN$_x$ and graphene, stronger bonding strength between TiN$_x$ and graphene, and/or inelastic phonon transmission across the TiN$_x$/G interfaces. To assess the role of mismatch of phonon energies, we first plot the compiled thermal conductance as a function of $\omega_{max}$, see

Figure 3a. We find that in general, the thermal conductance of graphene interfaces correlates well with the phonon energy in metals; metals with a high $\omega_{max}$ have a high thermal conductance. We compare the experimental thermal conductance to the calculations of the phonon radiation limit (eq 3) in Figure 3b. We find that for most metal/graphene interfaces (except Ti and TiN$_x$), $G_{exp} \approx 0.5 \times G_{rad}$. The fact that $G_{exp} \leq G_{rad}$ suggests that, unlike the metal/diamond interfaces[19-21], heat transport across most graphene interfaces is mainly due to elastic transmission of phonons. Since the transmission is elastic, the transmission probability of phonons from metals to graphene is limited by $\alpha_{rad}$ in eq 2 in Methods. $\alpha_{rad}$ and thus $G_{rad}$ could be optimized by matching phonon flux in graphene, $h_G(\omega)$, to phonon flux in metal, $h_M(\omega)$, over a wide range of phonon frequency. (The definition and discussion on $\alpha_{rad}$, $h_G(\omega)$ and $h_M(\omega)$ are given in Methods.) This can be achieved by ensuring that the phonon dispersions of graphene and metal are similar, see Figure S3 in Supplementary for the calculated $\alpha_{rad}$ for a few metal/graphene interfaces. We note that compared to other metals, $\alpha_{rad}$ for TiN$_x$/G interface shows two favorable characteristics that lead to higher thermal conductance of TiN$_x$/graphene interface: (1) $\alpha_{rad}$ is generally higher for low-frequency phonons and (2) more high-frequency phonons could transmit to graphene due to the higher $\omega_{max}$.

Next, we examine the role of interfacial binding strength by plotting the ratios of the thermal conductance measurements to the calculations of the corresponding phonon radiation limit, $G_{exp} / G_{rad}$, as a function of the interfacial binding energy $E_b$ of graphene interfaces in Figure 3c. In the figure, we approximate the binding energy $E_b$ from the absorption energy of graphene flakes absorbed on metal substrates derived from first principles calculations.[17] (We note that for $TiN_x$, we use the binding energy of Ti instead, as we are not able to find the first principles calculation for $TiN_x$.) We find that for most metals, no matter whether chemisorption (i.e., Pd, Ni) or physiosorption (i.e., Ag, Au, Al) bonds are formed with graphene, $G_{exp} / G_{rad}$ does not depend strongly on $E_b$ even when $E_b$ varies by a factor of 4, see Figure 3c.

Interesting, we find that, contrary to other metals, $G_{exp} / G_{rad}$ exceeds 1 for Ti and $TiN_x$. The fact that $G_{exp} / G_{rad} > 1$ suggests the existence of an additional channel for heat transport across the interfaces, supplementary to the elastic phonon transmission considered in the phonon radiation limit. One possible mechanism is sizeable inelastic heat transport across the interfaces, which is well accepted as the dominant heat transport across interfaces of diamonds.[20-21] The exact reasons for the larger inelastic phonon transmission across the Ti-carbon bonds are unknown to the authors.

In conclusion, we demonstrate a simple approach to enhance the phononic heat transport across metal/graphene interfaces, through a better match of phonon energy in the metals and graphene. The approach leads us to a huge thermal conductance for metallic $TiN_x$ contacts on graphene. Interestingly, the thermal conductance of $TiN_x$/graphene interface is larger than the phonon radiation limit, suggesting inelastic phonon transmission could partially contribute to heat transfer across the interface. Our work also advances the understanding on the mechanisms of heat transport across graphene interfaces.

## Methods

### Sample Preparation

For our samples, we choose GaN/sapphire with high thermal conductivity as the substrate to improve the accuracy of our thermal measurements.[13] We purchase the chemical vapor deposition (CVD) grown graphene on copper foils from Graphene Supermarket, and follow procedures stated in ref 10 to achieve clean graphene transfer. We choose poly(bisphenol A carbonate) (PC) instead of poly(methyl methacrylate) (PMMA) as the supporting layer during the graphene transfer, because it is easier to completely dissolve PC in chloroform.[33] The top metal films are ≈ 100 nm thick, while the bottom metal films are much thicker (300 – 400 nm) to reduce the sensitivity of the thermal measurements to the thermal conductance of the bottom metal/GaN interfaces, a source of uncertainty for our thermal measurements. The Au and Ag films in our samples were deposited by thermal evaporation at a rate of 10 Å/s with a base pressure of $10^{-8}$ Torr. The $TiN_x$ films were deposited by electron beam (e-beam) evaporation, and post-annealed at 923 K for 4 hours with a base pressure of 3 mTorr to increase the crystallinity of the films.

### Sample Characterization and TDTR Measurements

For the Raman spectroscopy measurements, we first deposited a thin metal (Au, Ag or $TiN_x$) film (5 – 10 nm) on graphene/$SiO_2$ by thermal or e-beam evaporation. We then measured the Raman spectra of the metal/graphene/$SiO_2$ samples using a home-built Raman system with a 532 nm wavelength continuous laser, see Figure 1c. We observe that the G peak of the samples are red shifted to 1606 – 1615 cm$^{-1}$, and from the magnitude of this shift, we estimate that a carrier concentration[34] of > 8 × 10$^{12}$ cm$^{-2}$ is induced in the graphene due to the charge transfer from the metals.

We quantify the conformity of graphene to the substrates from the AFM topographic images, see ref 10 for the details. To do so, we obtain the relative height $h$ from the depth histogram in Figure 1d, and derive the accumulative percentage of area $A(h)$ from the total area of the AFM topographic images with a relative height higher than $h$. We then plot $A(h)$ as $h$ for our samples before and after graphene transfer in figure 1f. The graphene is considered conformal if the difference in the relative height $\Delta h$ of before and after the transfer is ≤ 0.5 nm.

We conduct our TDTR measurements using a home-built setup; a schematic diagram of our setup is included in ref 25. For the samples coated with Au and TiN$_x$, the TDTR measurements were performed using a 5× objective lens with 1/e$^2$ radii of 10.2 μm and a laser power of < 600 mW, to limit the steady state temperature rise to ≈ 10 K. For the samples coated with Ag, however, we need to increase the steady state temperature rise to ≈ 20 K due to the relatively weak $dR/dT$ of Ag. We thus used a 10× objective lens with 1/e$^2$ radii of 5.1 μm and a total laser power of 400 mW. Following a method[35] reported earlier, we derived $dR/dT$ of Ag to be 3.6 × 10$^{-5}$ K$^{-1}$ at a laser wavelength of 780 nm.

We derive $G_{M/G/M}$ by comparing the ratio of in-phase and out-of-phase signals of TDTR measurements to calculations of a thermal model.[25, 36] In the analysis, $G_{M/G/M}$ is the only fitting parameter, and all other parameters are obtained either from literature or stand-alone measurements. We are not able to determine the thickness of the Au and Ag films by picosecond acoustics[37] as in other transducer films (e.g., TiN$_x$), because the acoustic echoes are too weak for the noble metals. Instead, we derive the thickness from AFM images of the thin films over sharp edges fabricated by photolithography. For Au and Ag films, we determine the thermal conductivities from the electrical resistivities measured by a four-point probe, using the Wiedemann-Franz law. For the TiN$_x$ film, however, we independently measured the thermal conductivity of TiN$_x$ by TDTR using a 330 nm TiN$_x$ film on SiO$_2$/Si, because the phononic thermal conductivity of TiN$_x$ is substantial. We derived 7.5 W m$^{-1}$ K$^{-1}$ at room temperature for our TiN$_x$ films. We also independently measured the thermal conductance of metal/GaN interfaces by TDTR to improve the accuracy of the measurements of $G_{M/G/M}$. With the aforementioned careful consideration, we manage to reduce the uncertainties of the derived $G_{M/G/M}$ to 7 – 40%.

## Calculations of the Radiation Limits

The phonon radiation limit of metal/graphene interfaces can be estimated from the properties of the metal ($C_j(\omega)$ and $v_j(\omega)$) and the transmission probability from the metal to graphene $\alpha_{rad}(\omega)$ using[23]

$$G_{rad} = \frac{1}{4}\sum_j \int_0^{\omega_{max,j}} C_j(\omega) v_j(\omega) \alpha_{rad}(\omega)\, d\omega \tag{1}$$

where $C_j(\omega) = \hbar\omega D_j(\omega) \frac{\partial f(\omega)}{\partial T}$ is the lattice heat capacity of the metal for phonon modes of frequency $\omega$ of phonon polarization $j$, $D_j(\omega) = \frac{\omega^2}{2\pi^2 v_j^3(\omega)}$ is the density of states for phonons mode of frequency $\omega$, $f$ is the Bose-Einstein distribution, $T$ is the temperature, $v_j(\omega)$ is the phonon group velocity of the metal, $\omega_{max,j}$ is the maximum phonon frequency of the respective phonon branch $j$ and $\alpha_{rad}(\omega)$ is given by

$$\alpha_{rad}(\omega) = \begin{cases} 1 & h_G(\omega) \geq h_M(\omega) \quad (2a) \\ h_G(\omega)/h_M(\omega) & h_G(\omega) < h_M(\omega) \quad (2b) \end{cases}$$

eq 2b is a result of the detailed balance of phonon transmission across the interfaces. Here, $h_G(\omega)$ and $h_M(\omega) = \sum_j v_j(\omega) D_j(\omega) / 4$ are phonon flux from the graphene side and the metal side, respectively. $h_G(\omega)$ and $h_M(\omega)$ are sometimes called vDOS in the literature,[38] since they are sums of products of the component of phonon velocity ($v_j$) normal to the interface (i.e., along c-axis in our cases) and density of phonon modes (DOS) $D_j(\omega)$.[38] In this work, we approximate $h_G(\omega)$ from properties of graphite using an anisotropic model that we recently developed; details of our calculations are presented below and in ref 39.

When $h_G(\omega) < h_M(\omega)$ for all phonons, which is a good approximation for many metals, we prove in Section 1 of Supplementary that the phonon radiation limit $G_{rad}$ only depends on the highest frequency (or equivalently the highest energy) of phonons, $\omega_{max}$, and not on other properties of metals,

$$G_{rad} = \int_0^{\omega_{max}} \hbar\omega \frac{\partial f(\omega)}{\partial T} h_G(\omega) \, d\omega \quad (3)$$

Note that $h_G(\omega)$ is the phonon flux in *graphite*, and eq 3 is an integration to the highest phonon frequency in the *metal*, $\omega_{max}$.

In the implementation, we approximate $h_G(\omega)$ using an anisotropic model with truncated linear dispersion of graphite; refer to ref 39 for the details of the anisotropic model. For Au, Ag, Pd, Al, Ni, Cr, we approximate $\omega_{max}$ from the average values of the maximum phonon frequencies of longitudinal acoustic (LA) phonons along [001], [011] and [111] directions. For Ti and TiN$_x$, we include optical phonons in the calculations since the velocity of optical phonons is considerable, and thus derive $\omega_{max}$ from the maximum frequencies of optical phonons. For amorphous SiO$_2$, we approximate $\omega_{max}$ from the Debye cutoff frequency of LA branch, $\omega_{max} = v_L(6\pi^2 n)^{\frac{1}{3}}$, where $v_L$ is the speed of sound of LA phonons, $n$ is the atom density of SiO$_2$. For $h_G(\omega)$, readers are referred to ref 39 for details.

## Acknowledgements

This work is supported by the Singapore Ministry of Education Academic Research Fund Tier 2, under Award No. MOE2013-T2-2-147 and Singapore Ministry of Education Academic Research Fund Tier 1 FRC Project FY2016. Sample characterization was carried out in part in the Centre for Advanced 2D Materials.

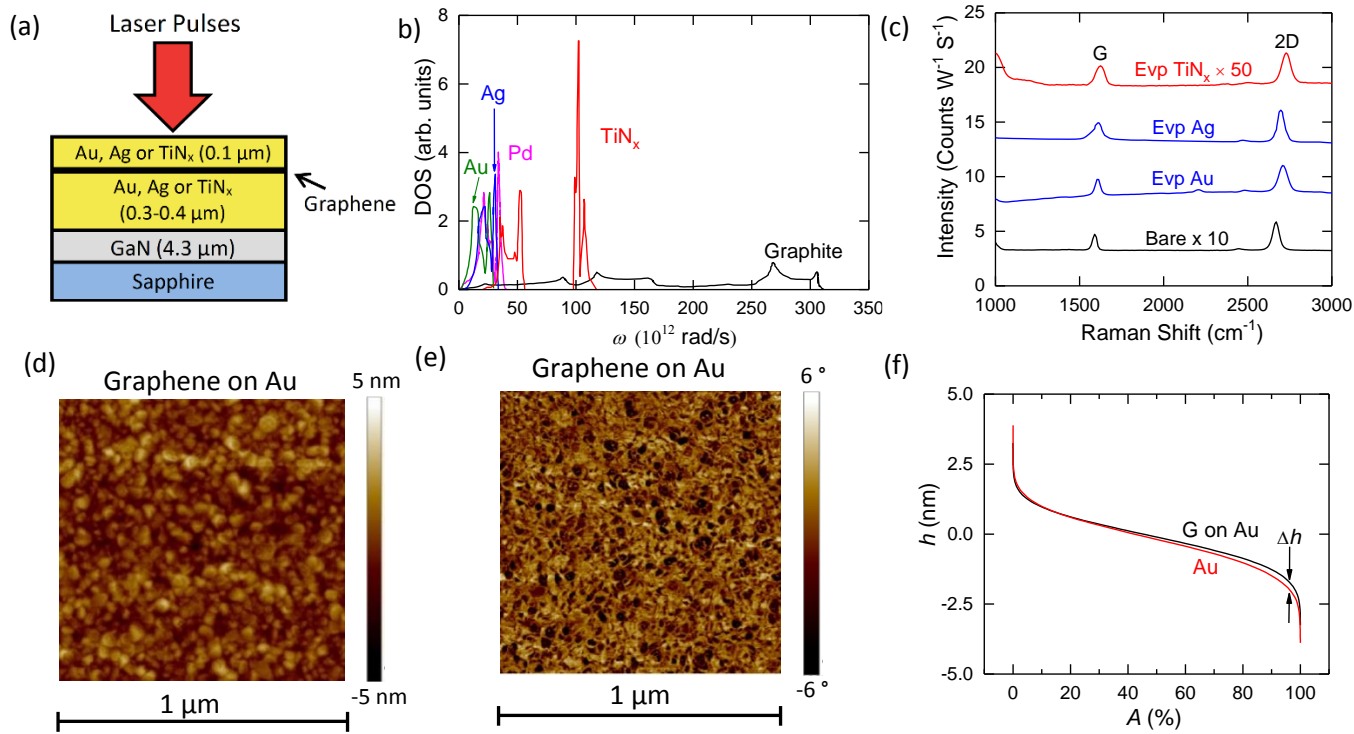

**Figure 1:** **(a)** Cross-section schematic diagram of our metal/graphene/metal samples. **(b)** Comparison of phonons density of states of graphite (black, ref 40), Au (green, ref 41), Ag (blue, ref 42), Pd (pink, ref 43) and TiN$_x$ (red, ref 44). **(c)** Raman spectra of CVD graphene transferred on SiO$_2$ coated with approximately 5 to 10 nm thick metal (i.e., TiN$_x$, Au or Ag) film, deposited either by thermal evaporation (blue) or by electron-beam evaporation (red), compared to that of a bare sample without any metal (black). The intensity of the Raman spectra of TiN$_x$ and bare samples are multiplied by a factor of fifty and ten, respectively, and all spectra are shifted vertically for ease of comparison. We find that the intensities of the graphene samples deposited with Au and Ag are significantly higher compared to that deposited with TiN$_x$, due to different enhancement factors by the metals in the surface enhanced Raman scattering (SERS) of graphene. **(d)** A representative topographic image of transferred graphene on Au, acquired by tapping mode atomic force microscopy (AFM). The rms roughness of the AFM image is 0.96

nm, similar to the 0.86 nm rms roughness of the bottom Au film. **(e)** phase-contrast image of the AFM measurement of (d), indicating no visible polymer residues. **(f)** Relative height of transferred graphene on Au and the Au substrate derived from AFM depth histogram in Figure 1(d), plotted as a function of accumulative percentage of area *A*. We assume the graphene is conformal when $\Delta h < 0.5$ nm.

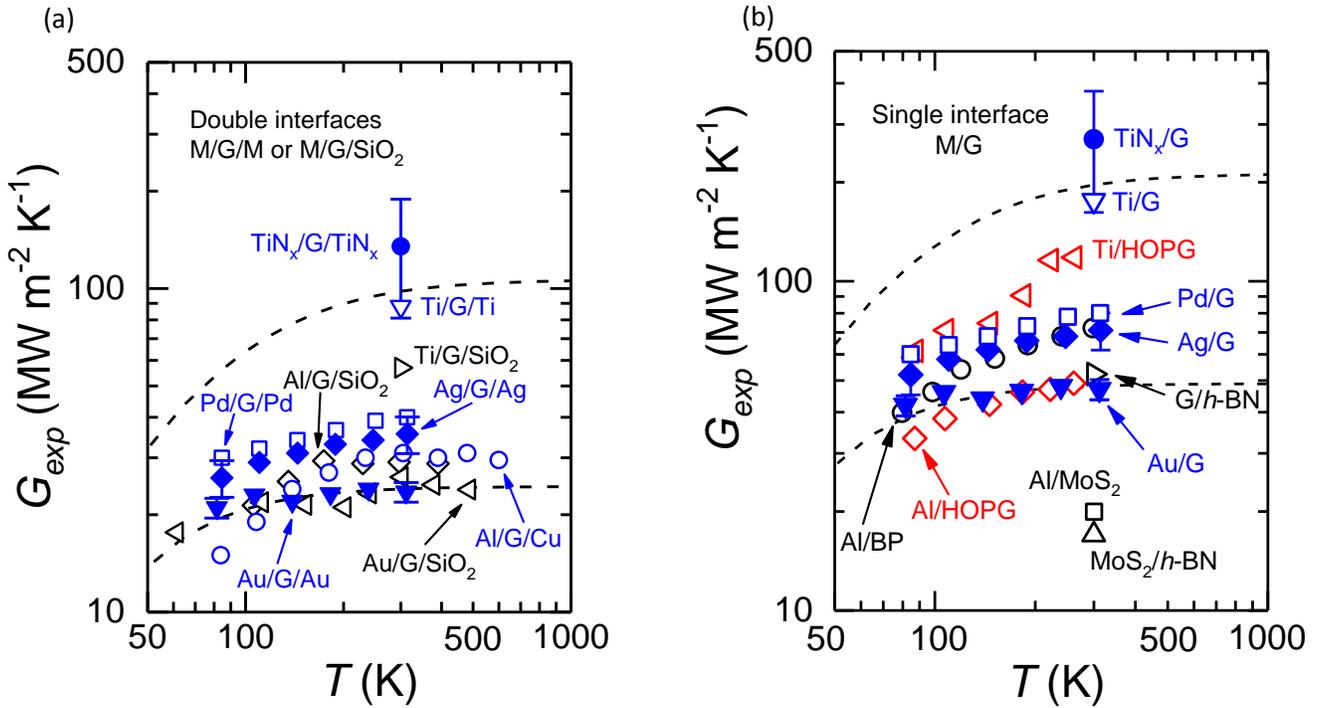

**Figure 2:** **(a)** Temperature dependence of the measured thermal conductance $G_{exp}$ of interfaces of $TiN_x/G/TiN_x$ (solid circles, this work), Ag/G/Ag (solid diamonds, this work) and Au/G/Au (solid down triangles, this work), compared to that of Al/G/Cu (open circles, ref 10), $Au/G/SiO_2$ (open left triangles, ref 7), $Al/G/SiO_2$ (open diamonds, ref 12), $Ti/G/SiO_2$ (open right triangle, ref 8), Pd/G/Pd (open squares, ref 13), Ti/G/Ti (open down triangle, ref 14). Blue symbols represent $G_{exp}$ of metal/graphene/metal interfaces, while black symbols represent $G_{exp}$ of metal/graphene/$SiO_2$ interfaces. The dashed lines are calculations of $G_{rad}$ of $TiN_x/G/TiN_x$ (top) and $0.5 \times G_{rad}$ of Au/G/Au (bottom), assuming that the thermal resistance of metals/graphite and graphite/metals interfaces adds in series. **(b)** Temperature dependence of the thermal conductance $G_{exp}$ of single interfaces of $TiN_x/G$ (solid circles, this work), Ag/G (solid diamonds, this work) and Au/G (solid down triangles, this work), Pd/G (open squares, ref 13), Ti/G (open down triangle, ref 14), derived from measurements in **(a)**, compared to that of Ti/HOPG (open left triangles, ref 26), Al/BP (open circles, ref 27), Al/HOPG (open diamonds, ref 26), $MoS_2$/h-

BN (open up triangle, ref 29), G/h-BN (open right triangle, ref 29) and Al/MoS$_2$ (open square, ref 28). Blue symbols represent $G_{exp}$ of single M/G interfaces derived from M/G/M interfaces, black symbols represent $G_{exp}$ of other 2D materials interfaces, red symbols represent $G_{exp}$ of metal/HOPG interfaces. The dashed lines are calculations of $G_{rad}$ of TiN$_x$/G/TiN$_x$ (top) and 0.5×$G_{rad}$ of Au/G/Au (bottom).

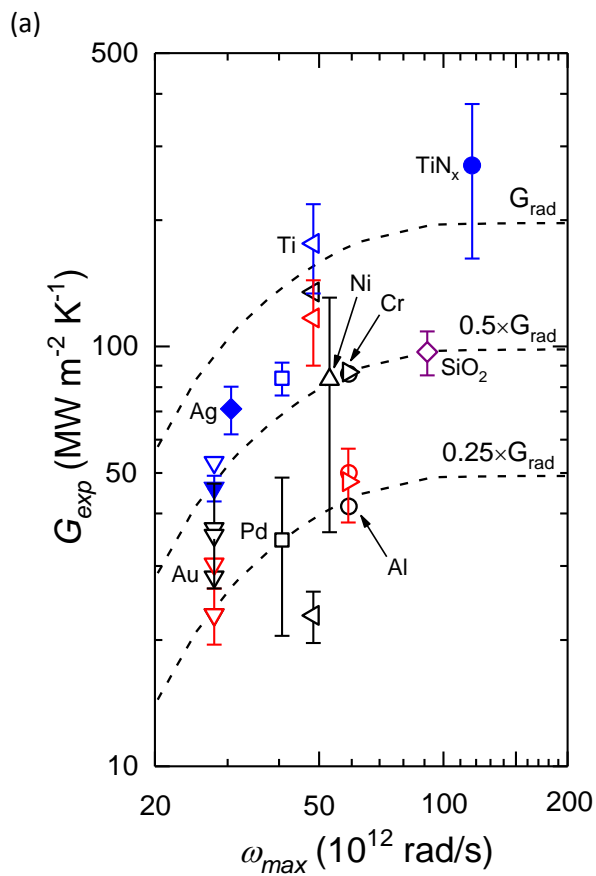
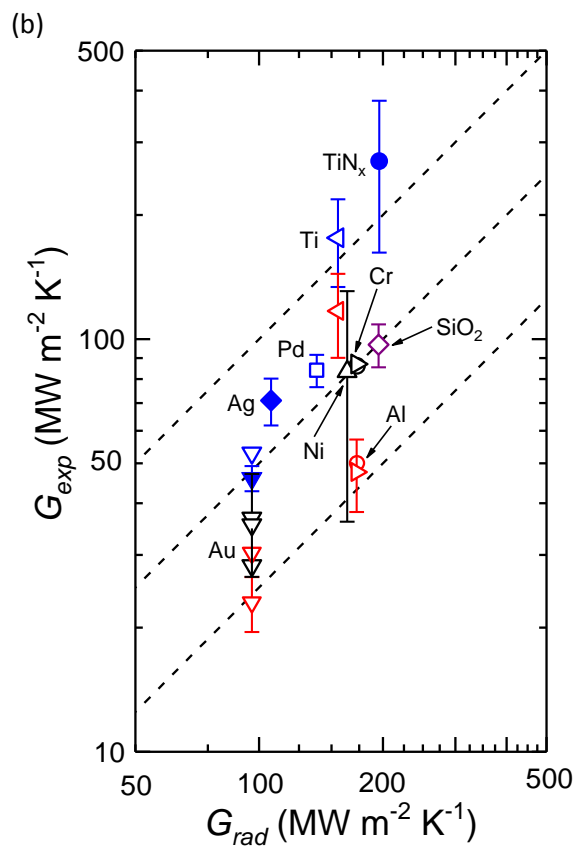
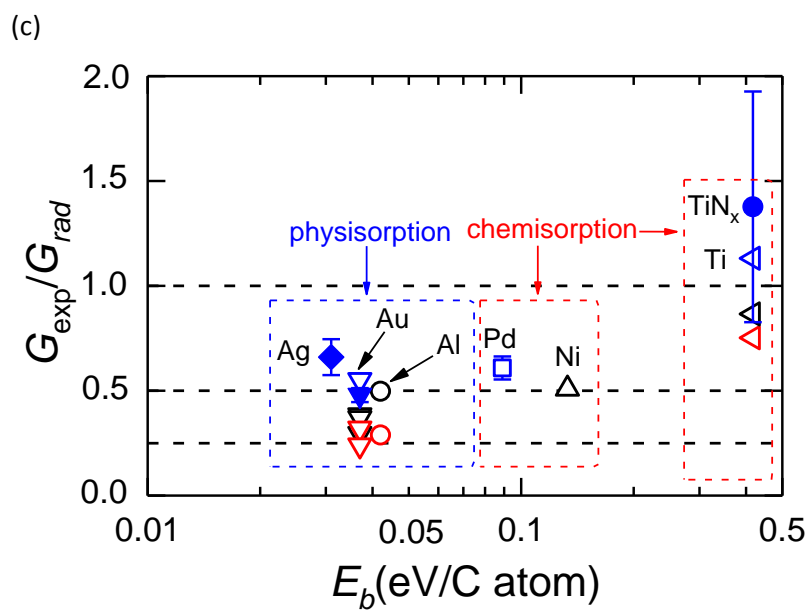

**Figure 3: (a)** Compilation of thermal conductance $G_{exp}$, as a function of the highest frequency of phonons of metals (or SiO$_2$) $\omega_{max}$. The metal contacts that we compile include gold (down triangle, refs 7, 14, 16, 26, 30), silver (diamond), palladium (square, refs 8, 13), aluminum (circle, refs 8, 12, 26), titanium nitride (circle), titanium (left triangle, refs 8, 14, 26, 30), nickel (up triangle, ref 8) and chromium (right triangle, refs 8, 26), silicon dioxide (diamond, ref 22), as labeled. Solid symbols represent measurements in this work, while open symbols represent measurements from literature. Blue symbols represent $G_{exp}$ of single M/G interfaces derived from M/G/M interfaces, black symbols represent $G_{exp}$ of single M/G interfaces derived from M/G/SiO$_2$ interfaces, red symbols represent $G_{exp}$ of metal/HOPG interfaces, and the purple symbol represents $G_{exp}$ of graphene/SiO$_2$ interface. We reestimate the uncertainties of some measurements from refs 8 and 22. **(b)** Comparison of the experimental thermal conductance ($G_{exp}$) in (a) to the calculations of the phonon radiation limit ($G_{rad}$). We use the same symbols as in (a). Some measurements in (a) are omitted for clarity. **(c)** Compilation of ratios of $G_{exp}$ / $G_{rad}$, as a function of binding energy between the metals and graphene $E_b$.[17] Au, Ag, Al form physisorption bonding with graphene, while Pd, Ni, Ti and TiN$_x$ form chemisorption bonding, as indicated. The dashed lines in all (a)-(c) are calculations of (from the top) $G_{rad}$, 0.5×$G_{rad}$ and 0.25×$G_{rad}$.

# Supporting Information

# Achieving huge thermal conductance of metallic nitride on graphene through enhanced elastic and inelastic phonon transmission

Weidong Zheng[1,2], Bin Huang[1,2], Hongkun Li[1], Yee Kan Koh[1,2*]

[1]Department of Mechanical Engineering, National University of Singapore, Singapore 117576

[2]Centre for Advanced 2D Materials, National University of Singapore, Singapore 117542

*Corresponding Author. Email: mpekyk@nus.edu.sg

# S1: Derivation of radiation limit $G_{rad}$ of metal/graphene interface

According to eq 1,

$$G_{rad} = \frac{1}{4}\sum_j \int_0^{\omega_{max,j}} \hbar\omega D_j(\omega) \frac{\partial f(\omega)}{\partial T} v_j(\omega)\alpha_{rad}(\omega)d\omega \quad (S1)$$

We regard the group velocities $v_j(\omega)$ as constant, $v_L$ and $v_T$ for longitudinal acoustical (LA) and transverse acoustical (TA) branches, respectively, and the density of states for LA and TA branches are,

$$D_L(\omega) = \frac{\omega^2}{2\pi^2 v_L^3} \quad (S2)$$

$$D_T(\omega) = \frac{\omega^2}{2\pi^2 v_T^3} \quad (S3)$$

Considering there are two branches for TA and one branch for LA, when $h_G(\omega) < h_M(\omega)$,

$$G_{rad} = \frac{1}{4}\int_0^{\omega_{max,L}} \hbar\omega \frac{\partial f(\omega)}{\partial T} \frac{\omega^2}{2\pi^2 v_L^2} \frac{h_G(\omega)}{h_M(\omega)} d\omega + \frac{1}{2}\int_0^{\omega_{max,T}} \hbar\omega \frac{\partial f(\omega)}{\partial T} \frac{\omega^2}{2\pi^2 v_T^2} \frac{h_G(\omega)}{h_M(\omega)} d\omega \quad (S4)$$

In general, $\omega_{max,L} > \omega_{max,T}$,

$$G_{rad} = \frac{1}{4}\int_0^{\omega_{max,T}} \hbar\omega \frac{\partial f(\omega)}{\partial T} \frac{\omega^2}{2\pi^2 v_L^2} \frac{h_G(\omega)}{h_M(\omega)} d\omega + \frac{1}{4}\int_{\omega_{max,T}}^{\omega_{max,L}} \hbar\omega \frac{\partial f(\omega)}{\partial T} \frac{\omega^2}{2\pi^2 v_L^2} \frac{h_G(\omega)}{h_M(\omega)} d\omega$$
$$+ \frac{1}{2}\int_0^{\omega_{max,T}} \hbar\omega \frac{\partial f(\omega)}{\partial T} \frac{\omega^2}{2\pi^2 v_T^2} \frac{h_G(\omega)}{h_M(\omega)} d\omega \quad (S5)$$

The phonon flux from the metal side is[1],

$$h_M(\omega) = \begin{cases} 0 & \omega_{max,L} < \omega \\ \dfrac{\omega^2}{8\pi^2 v_L^2} & \omega_{max,T} \leq \omega \leq \omega_{max,L} \\ \dfrac{\omega^2}{8\pi^2 v_L^2} + \dfrac{\omega^2}{4\pi^2 v_T^2} & 0 \leq \omega < \omega_{max,T} \end{cases} \quad (S6)$$

Substitute eq S6 into eq S5, we get the radiation limit of metal/graphene interface,

$$\begin{aligned} G_{rad} &= \int_0^{\omega_{max,T}} \hbar\omega \frac{\partial f(\omega)}{\partial T} h_G(\omega) \left( \frac{1}{1 + 2v_L^2/v_T^2} + \frac{1}{1 + v_T^2/(2v_L^2)} \right) d\omega + \int_{\omega_{max,T}}^{\omega_{max,L}} \hbar\omega \frac{\partial f(\omega)}{\partial T} h_G(\omega) d\omega \\ &= \int_0^{\omega_{max,L}} \hbar\omega \frac{\partial f(\omega)}{\partial T} h_G(\omega) d\omega \end{aligned} \quad (S7)$$

Therefore, the phonon radiation limit $G_{rad}$ only depends on the highest frequency of phonons, $\omega_{max,L}$, and not on other properties of metals.

## S2: Thermal conductance comparison of approximation and full calculations

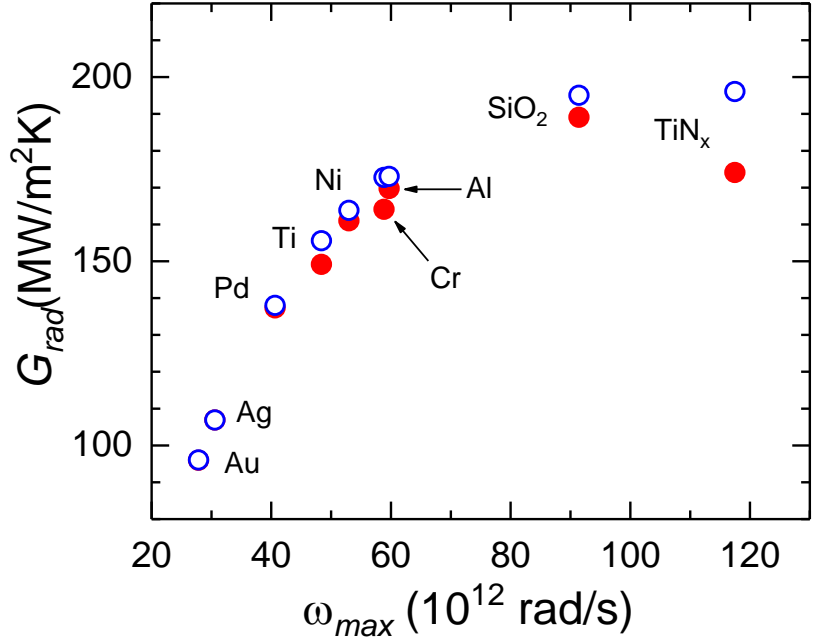

**Figure S2:** Comparison of two radiation limit calculation cases for all the metal/graphene interfaces we consider in this work. Case 1 (blue open circles), approximation calculation, transmission probability $\alpha_{rad}$ is determined with eq 2b for the all phonons. Case 2 (red solid circles), full calculation, transmission probability is restricted to $\alpha_{rad} \leq 1$ by both eq 2a and eq 2b.

When $h_G(\omega)/h_M(\omega) > 1$, eq 3 cannot be used to determine the radiation limit, $G_{rad}$ of metal graphene interface. However, Figures S2 shows that error of these two cases is generally less than 10 % for all the metal/graphene interfaces we consider in this paper. The number of phonons with $h_G(\omega) < h_M(\omega)$ is negligible compared to the total amount of phonons. Therefore, we employ eq 3 for all our analysis in this work.

## S3: Comparison of transmission probability from metals to graphene

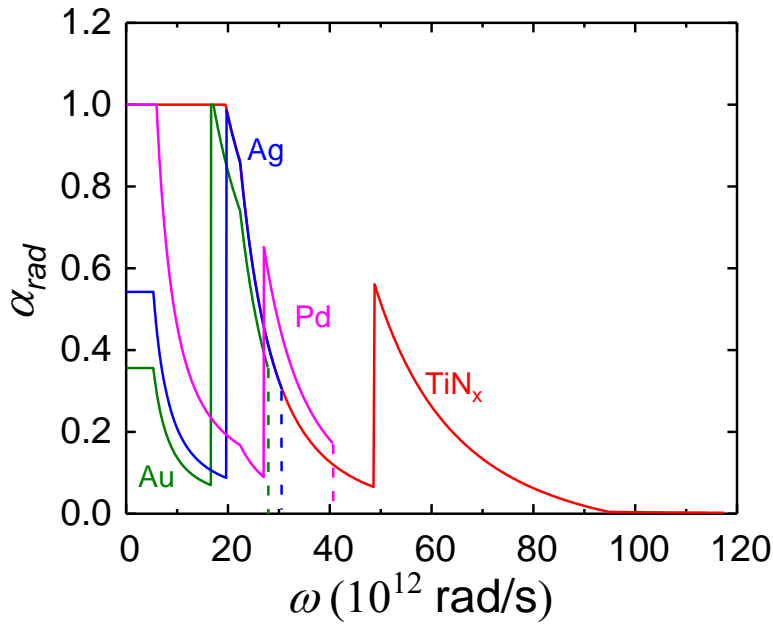

**Figure S3:** Comparison of transmission probability from metals to graphene according to eq 2a and eq 2b, Pd (pink), Au (green), Ag (blue) and $TiN_x$ (red). The transmission probability for $TiN_x$/graphene interface shows two characteristics: (1) $\alpha_{rad}$ is generally higher for low-frequency phonons; (2) more high-frequency phonons could transmit to graphene. These two characteristics could contribute to higher thermal conductance of $TiN_x$/graphene compared to other metal/graphene interfaces.